\documentclass[superscriptaddress, reprint, amsmath, amssymb, aps, prl]{revtex4-1}
\normalsize {\tiny }
\usepackage{graphicx}% Include figure files
\usepackage{dcolumn}% Align table columns on decimal point
\usepackage{bm}% bold math
\usepackage{color}
\usepackage{epstopdf}

\usepackage{listings}
\begin{document}

\title{An Alternative Strategy to Control the Electronic Properties of Bilayer Graphene: Semi-metal to Metal Transition and a 2D Material with Dirac Cone}

\author{Srimanta Pakhira}
\affiliation{Condensed Matter Theory, National High Magnetic Field Laboratory, Scientific Computing Department, Materials Science and Engineering, Florida State University (FSU), Tallahassee, Florida, 32310, USA}
\affiliation{Department of Chemical \& Biomedical Engineering, FAMU-FSU Joint College of Engineering, Tallahassee, Florida, 32310, USA.}
\author{Kevin P. Lucht}%
\author{Jose L. Mendoza-Cortes}
\email{Corresponding Author: Prof. J. L. Mendoza-Cortes, email: mendoza@eng.famu.fsu.edu\\[-3pt]} 

\affiliation{Condensed Matter Theory, National High Magnetic Field Laboratory, Scientific Computing Department, Materials Science and Engineering, Florida State University (FSU), Tallahassee, Florida, 32310, USA}
\affiliation{Department of Chemical \& Biomedical Engineering, FAMU-FSU Joint College of Engineering, Tallahassee, Florida, 32310, USA.}

\date{\today}% It is always \today, today,

\begin{abstract}
The AB-stacked bilayer graphene (BLG) is a pure semiconductor whose band gap and properties can be tuned by various methods such as doping or applying gate voltage. Here we show an alternative method to control the electronic properties of BLG; intercalation of transition metal atoms between the two monolayers graphene (MLG). A theoretical investigation has been performed on two-dimensional MLG, BLG, BLG-intercalated nanostructured materials, all of which are energetically stable. Our study reveals that only MLG and BLG-intercalated with one Vanadium (V) atom (BLG-1V) materials have a Dirac Cone at the K-point. This study reveals a new strategy to control the material properties of BLG to exhibit various behaviors, including: metallic, semi-metallic, and semiconducting by varying the concentration and spin arrangement of the V atoms in BLG. In all the cases, the present DFT calculations show that the 2p$_z$ sub-shells of C atoms in graphene and the 3d$_{yz}$ sub-shells of the V atoms provide the electron  density near the Fermi level controlling the material properties of the BLG-intercalated materials. Thus we prove that out-of-plane atoms can influence in plane electronic densities in BLG.
\end{abstract}
\maketitle

%\tableofcontents
%\section{\label{sec:level1}INTRODUCTION}
In modern science and technology, there has been a tremendous amount of theoretical and experimental interest in the low energy electronic properties of ultrathin graphite films including graphene monolayer and graphene bilayer materials \cite{Novoselov2005}. Graphene \cite{Novoselov2004}, a 2D honeycomb sheet of carbon just one atom thick with hybridized sp$^2$ bonded orbitals between carbon atoms \cite{CastroNeto2009}, is an ideal and novel material for making nanoelectronic and photonic devices, because it is a very good electrical conductor as well as being the thinnest 2D material known. Graphene has a unique linear band structure around the Fermi level (E$_F$) forming a Dirac Cone at the K-points of its Brillouin zone, and has led to fascinating phenomena, exemplified by massless Dirac fermion physics \cite{Ohta2006, Castro2007, Zhou2007, Mullen2015}. This emergent behavior of Dirac fermions in condensed matter systems defines the unifying framework for a class of materials called Dirac materials.

Monolayer graphene (MLG) or graphene has an electronic structure that can be controlled by an electrical field \cite{Novoselov2012}. To be used in digital electronics, however, MLG has the well known zero-gap issue \cite{Avouris2010} which makes a high on-off ratio difficult; deeming it unsuitable for transistors, which are the foundation of all modern electronic devices. Bilayer graphene (BLG) can be used instead of MLG to overcome the zero-gap problem, with a gap opening simply by applying an electric field \cite{Oostinga2008, McCann2007}. BLG has an entirely different band structure, and equally interesting \cite{Ohta2007, Nath2015}, since the band gap of the BLG can be modulated from zero to a few eV by using different methods such as doping and applying an external electric field. In addition, BLG holds the potential for electronics and nano-technological applications, particularly because of the possibility to control both carrier density and energy band gap through doping or gating \cite{McCann2006, Castro2007, Novoselov2006, Ohta2006, Nath2015}. The most remarkable property of BLG is that the inversion symmetric AB-stacked BLG is a zero-band gap semiconductor in its pristine form, but a non-zero band gap can be induced by breaking the inversion symmetry of the two graphene monolayers to form AA-stacked BLG. When two graphene monolayers are stacked (in both the cases AA- and AB-stacked), the monolayer features are lost and dispersion effects become quadratic and more effective \cite{McCann2006}. Thus, BLG acts as a semiconductor and exhibits an induced electric field and broadly tunable band gap \cite{Castro2007}.  Because tuning the band gap of BLG can turn it from a semiconductor into a metal or semi-metal, a single millimeter-square sheet of BLG could potentially hold millions of differently tuned electronic devices that can be reconfigured. Recently, lasers have been used to get BLG to act as a conductor, a Dirac material or semiconductor; an important step towards computer chips made of a 2D material \cite{Lin2015}. 

In this Letter we show that the band structure of BLG can be controlled by adding Vanadium (V) metal atoms between two graphene layers so that the electronic band gap between the valence and conduction bands can be tuned; thus resulting in the appearance of a Dirac Cone. By definition; intercalation occurs when one layer of metal atoms is inserted between two graphene monolayers. Intercalation of the V atoms between a graphene bilayer causes surprisingly a diverse array of electronic properties. We also show that the band gap depends on the coupling between the two graphene layers and symmetry of the BLG system i.e. AB stacking vs AA stacking. We have performed a theoretical investigation of the electronic properties such as band structure, projected density of states (DOSs), spin arrangements, and structural stability of the MLG, BLG, BLG intercalated with one Vanadium atom (BLG-1V), BLG intercalated with two Vanadium atoms (BLG-2V), and BLG intercalated with three Vanadium atoms (BLG-3V) in one unit cell. We have also investigated their structural stabilities by  computing the Gibbs free energies ($\Delta G\!_{f}$) of the aforementioned systems. 

%\onecolumngrid

\begin{figure}[h!]
	\centering
	\includegraphics[width=0.99\linewidth]{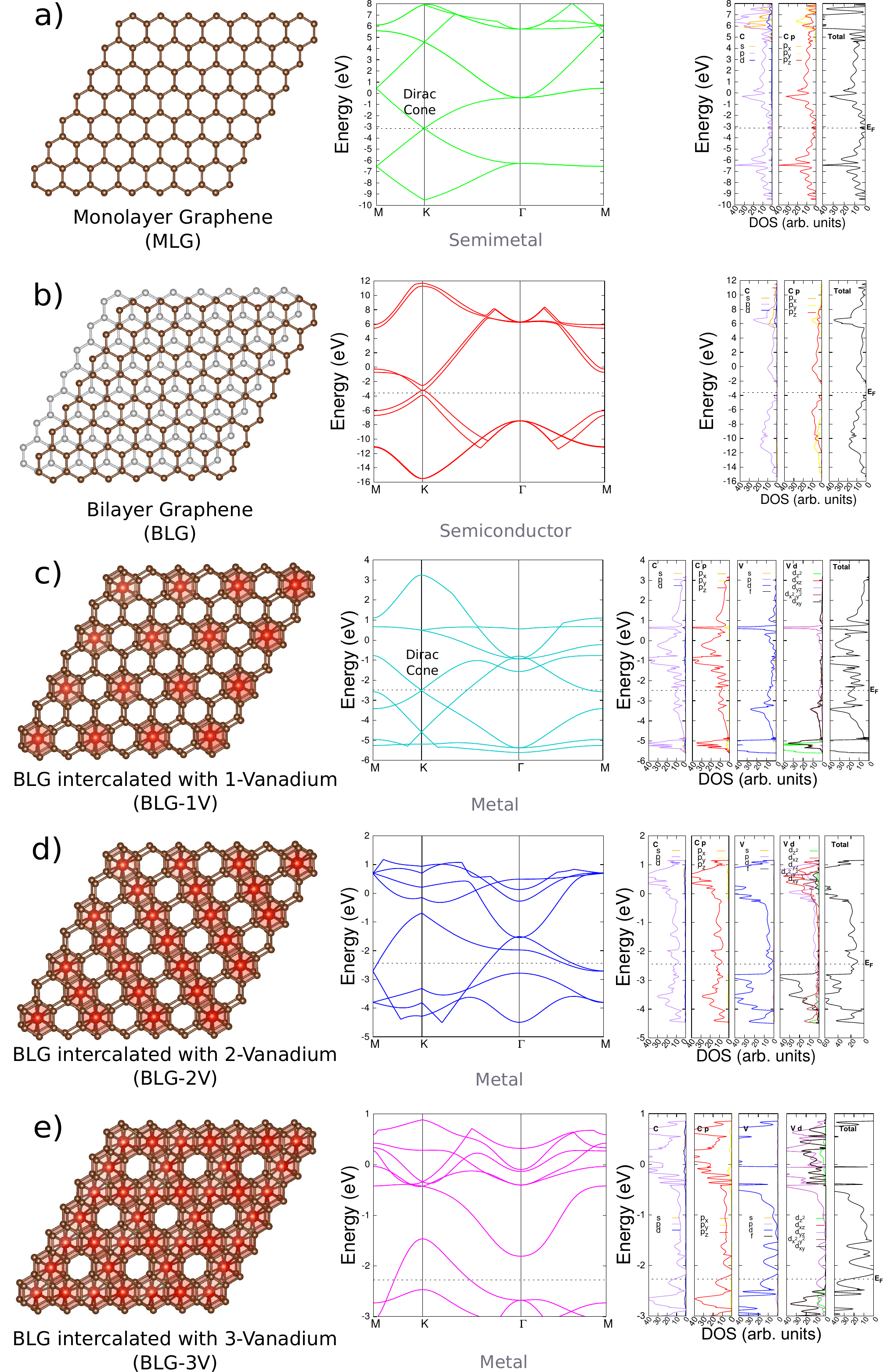}
	\caption{The optimized structures, band structures and density of states (DOSs) are shown a) MLG, b) BLG, c) BLG-1V, d) BLG-2V, and e) BLG-3V 2D materials. The individual components of the DOSs of the C and V atoms and total DOSs (depicted by ``Total'') are also presented in here.}
	\label{fig:V_graphene2DPROP}
	\vspace{-4mm}
\end{figure}

\twocolumngrid 
 
The geometry and 2D layer structure of the MLG, BLG and BLG-intercalated materials (BLG-1V, BLG-2V and BLG-3V) were optimized by dispersion-corrected hybrid density functional theory \cite{Becke1993, Becke1988, Lee1988, Grimme2006, Pakhira2012, Pakhira2013}, i.e. B3LYP-D2, which has been shown to give correct electronic properties of the 2D materials \cite{Grimme2006, Lucht2015}. The semi-empirical Grimme-D2 dispersion corrections were added in the present calculations in order to incorporate van der Waals dispersion effects on the system and to estimate the van der Waals forces \cite{Grimme2006, Pakhira2012, Pakhira2013}.  The CRYSTAL14 \cite{Dovesi2014} suite code was used to perform all the computations. The graphene intercalated nanostructured 2D materials (BLG-1V, BLG-2V, and BLG-3V) have been prepared by adding the V atoms in one unit cell of the crystal structures. In the present computation, triple-zeta valence with polarization quality (TZVP) Gaussian basis sets were used for both C and V atoms \cite{Peintinger2013}. Integration inside of the first Brillouin zone were sampled on 15 x 15 x 1 k-mesh grids for all the materials for both the optimization and material properties (band structures and density of states) calculations. We have plotted the bands along a high symmetric k-direction, $\mathrm{M-K-\Gamma-M}$, in the first Brillouin zone.  Electrostatic potential calculations have been included in the present computation i.e. the energy is reported with respect to the vacuum. In this work, both AB-stacked BLG system, and AA-stacked BLG-intercalated materials  have been considered with tunable interlayer separation. The unit cell constructed this way contains one configuration, known as AA, where one atom is exactly above another atom of the other layer of graphene (see Figure \ref{fig SM1:AA-Stacked-Structure} in the Supplementary Material), there is another configuration, AB, where an atom of the top of graphene layer is exactly in the center of a lower layer hexagon as shown in Figure 1b. To reproduce previous experimental and theoretical results, we have considered both AA and AB-stacking for the BLG material, however we have used the AA-stacked layer structures for the BLG-1V, BLG-2V, and BLG-3V materials, which are the stable configurations. The optimized geometries of the MLG, AB-stacked BLG, and BLG-intercalated materials along with their band structures and projected density of states (DOSs) are shown in Figure 1. The optimized geometry, band structures and DOSs of pristine AA-stacked BLG are reported in the Supplemental Material. The contribution of the sub-shells (such as p$_x$, p$_y$, p$_z$, d$_{yz}$ etc.) in the total DOSs has been also computed.  

The present DFT calculation shows that the Dirac Cone exists at the K-point in the band structure of the pristine MLG material which is consistent with previous results \cite{Ohta2006, Castro2007, Zhou2007, Mullen2015}. It is found that the valence bands of MLG arise from the 2p$_z$ sub-shells of C atoms, as depicted in the DOSs calculation. These sub-shells make the Dirac Cone in the band structure of the pristine MLG (as shown in Figure 1a). The total DOSs calculations show the electron density around the Fermi energy level (E$_F$), confirming MLG as a semimetal. 

For the pristine AB-stacked BLG material, the Fermi level lies below the point where the valence and conduction bands touch each other as shown in Figure 1b. We have also studied the AA-stacked BLG, but our computation shows that AB-stacked BLG is thermodynamically more stable than the AA-stacked BLG. The valence and conduction bands are crossing the E$_F$ at K-points for both the AA- and AB-stacked BLG materials. Thus, both the AA- and AB-stacked BLG materials have no Dirac Cone, which agree well with the previous experimental results \cite{Min2007, Latil2006}. Thus the AA-/AB-stacked BLG materials are ordinary non-zero/zero band gap semiconductor. The bond distances, lattice constants ($a$ and $b$), and intercalation distance ($d$) are reported in Table I. The distance between the two monolayers in the pristine AB-stacked BLG material is about 3.390 \AA\ which agrees with the previous experimental and theoretical results \cite{Min2007}. We have also calculated the binding energy ($\Delta G\!_{f}$) of all the system studied in here as shown in Table II. The required binding energy, $\Delta G\!_{f}$, to form the AB-stacked BLG from MLG, is about -0.86 eV. 

\begin{table}[h!]
	\vspace{-3mm}
	\caption{ Bond distances and lattice constants of the MLG, BLG, BLG-1V, BLG-2V, and BLG-3V 2D materials. The $d$ is defined as the interlayer distance between the monolayers.} 
	\label{table:mytable1}
	\tabcolsep=5.5pt
	\begin{tabular}{ c c c c c c }
		\hline 
		Component   &  C-C (\AA)  &  C-V (\AA)   &  $a$ (\AA) &  $b$ (\AA) & $d$(\AA) \\ 
		\hline 
		MLG         &  1.416   	  &    N/A 		 &  2.451     &  2.451     &  N/A      \\
		BLG         &  1.421      &    N/A  	 &  2.449     &  2.449     &  3.390     \\ 
		BLG-1V      &  1.439	  &   2.243	     &  4.942     &  4.942     &  3.441     \\ 
		BLG-2V      &  1.442	  &   2.335   	 &  8.575     &  5.022     &  3.645     \\
		BLG-3V      &  1.464      &   2.413      &  5.032     &  5.032     &  3.658      \\ 
		\hline 
		\vspace{-6mm}
	\end{tabular} 
\end{table}

\begin{table}[h!]
    \vspace{-4mm}
	\caption{Relevant properties of the MLG, BLG, BLG-1V, BLG-2V, and BLG-3V 2D materials. }
	\label{table:mytable2}
	\tabcolsep=6.5pt
	\begin{tabular}{ c c c c }
		\hline 
		Component  & State          & $\Delta G\!_{f}$ (eV) & Dirac Cone  \\ 
		\hline 
		MLG        & Semimetal      &   N/A                 &  Yes      \\
		BLG        & Semiconductor  &  -0.86                &  No	      \\ 
		BLG-1V     & Metal		    &  -5.97                &  Yes      \\ 
		BLG-2V     & Metal          &  -5.67                &  No         \\
		BLG-3V     & Metal          &  -5.36                &  No         \\ 
		\hline 
		\vspace{-7mm}
	\end{tabular} 
\end{table} 

The optimized structure between Vanadium and graphene follows an AA-stacking arrangement with the Vanadium placed at the center of the honeycomb, and it forms the BLG-1V 2D material. On this conformation, the Vanadium d-orbitals will be situated to favorably interact with the p$_z$ orbital orthogonal of the graphene layer. The BLG-1V structure is highly favorable by -5.97 eV relative to BLG. Additional Vanadium atoms are also favorable by -5.67 eV between single metal and double metal addition to form BLG-2V material, and by -5.36 eV  between the double metal and triple metal addition to form BLG-3V material as shown in Table II. The present DFT calculations show that the intercalation distances have been gradually increased in the intercalated BLG materials, BLG-1V to BLG-3V, but not equally as shown in Table I. Our computation also shows the C-C bond distance has been increased by an average of 0.014 \AA\ in the BLG-1V, BLG-2V and BLG-3V 2D materials due to the presence of V atoms between the graphene layers compared to BLG. We have calculated the stability of BLG material with four V atoms (BLG-4V), but the frequency calculations show that the structure is unstable thermodynamically as it has many imaginary frequencies, and hence this result is excluded from this Letter. 

The addition of Vanadium atom was found to substantially alter the electronic properties in the graphene bilayer. Addition of a single Vanadium atom intercalated in AA-stacked BLG (i.e. BLG-1V) yields a Dirac Cone along the K direction as shown in Figure 1c.  The DOSs around the Fermi energy level indicates metallic behavior. The individual components of the p-orbitals electron density of C atoms in graphene and the d-orbitals electron density of V atoms have been explicitly shown along with the total electron density in the DOSs computations as depicted in Figure 1c. The density projections show the electrons can freely move from valence band to the conduction band, and it reveals the intrinsic electron mobility of the BLG-1V materials. The DOSs calculations of the BLG-1V material show the electron density around the E$_F$ is coming from the p$_z$ (highlighted by red color) sub-shells of C atoms in graphene and the d$_{yz}$ (highlighted by violet color) sub-shell of V atom. In other words, the p$_z$ sub-shell of the carbon, and d$_{yz}$ sub-shells of Vanadium is solely responsible for the emergence of Dirac Cone of the BLG-1V material. In MLG, the Dirac Cone seems to come from p-orbital of C atoms (may be the other related materials as well) but for BLG-1V, the nature of the Dirac Cone is different because it comes from either p-orbital alone or d-orbital alone or a hybrid of p-d-orbitals at the same time. We can not tell for now but the difference in nature is interesting. As an excellent conductor and Dirac material, the BLG-1V material can be used in various modern electronic and nanotechnology devices.

BLG-2V can be formed with the addition of another Vanadium atom in BLG-1V material. This addition results in the Dirac Cone disappearing, and the bands to shift, showing a metallic behavior. The most stable spin states are reported in this Letter. All other spin states and the discussion about the spin alignment of the BLG-2V and BLG-3V 2D materials are described in the Supplementary Material. The 3d orbitals of V receive more electron donation from the graphene 2p$_z$ sub-shells when the total spin is increased.  The band structure reveals that along the M to K directions, the valence band crosses the Fermi level in both the BLG-2V and BLG-3V materials resulting in large electron distribution around the Fermi level as shown in Figure 1d and Figure 1e, respectively. Thus both BLG-2V and BLG-3V materials show metallic behavior as depicted in their DOSs. In both materials, the DOSs show that electron distribution or density around the Fermi level is due to the p$_z$ sub-shell electrons of C atoms in graphene and the d$_{yz}$ sub-shell electrons of V atoms as depicted in sub-shells DOSs calculations. Among all the intercalated materials, BLG-3V shows the least electron density near E$_F$.
%\vspace{-4.55mm}

We have studied the structure and material properties of the MLG, BLG, BLG-intercalated nanostructured 2D materials. The individual components of the sub-shells of the p-orbital of C atoms and the d-orbital of the V atoms, which are taking part in the total electron density in DOSs are also reported along with the total DOSs. Among all the systems, MLG and BLG-1V materials have a Dirac Cone at the K-point. Once the concentration of Vanadium is increased the Dirac Cone disappears. This work provides the first theoretical investigation on the BLG with intercalation of one, two and three V atoms per unit cell and the first observation of Dirac Cone in the BLG-1V material. It further shows how the material properties have been changed in BLG-1V, BLG-2V and BLG-3V due to the presence of Vanadium atoms. We have also found that the 2p$_z$ sub-shell of C atoms and 3d$_{yz}$ sub-shell of the V atoms in the BLG-1V, BLG-2V, and BLG-3V materials are the main components around E$_F$ playing the main role to show the Dirac Cone and the conducting properties. In conclusion, the Dirac Cone structure gives the MLG and BLG-1V materials massless fermions leading to ultrahigh carrier mobility, with the former having p-orbitals character while the second involves p- and d-orbitals character. 

%We can also conclude that BLG-1V, a Dirac material offers us a guidance in predicting the new features that we have shown to be universal, just like the currently well known Dirac materials which play an important role in modern material science. This study may allow the BLG, the only known semiconductor with a tunable energy gap, new avenues to develop 2D Dirac materials and modern electronic devices. 
     
\vspace{-5mm}

\bibliography{PRL-Bibliography}
\clearpage
\appendix

\newpage
\AtEndDocument{\clearpage}

%\AtEndDocument{\clearpage}

%%%%%%%%%% Merge with supplemental materials %%%%%%%%%%
\pagebreak
\widetext
\clearpage
%\documentclass[superscriptaddress, reprint, amsmath, amssymb, aps, prl]{revtex4-1}
%\normalsize {\tiny }
%\usepackage{graphicx}% Include figure files
%\usepackage{dcolumn}% Align table columns on decimal point
%\usepackage{bm}% bold math
%\usepackage{color}
%%\usepackage[version=3]{mhchem} % Formula subscripts using \ce{}
%
%\usepackage{listings}
%\lstset{
%	basicstyle=\scriptsize \ttfamily,
%	columns=fullflexible,
%	keepspaces=true,
%}
%
%\begin{document}
	
%\onecolumngrid

\begin{center}
	{\Huge{\textbf{Supplementary Material}}}
\end{center}
\vspace{13pt}
\setcounter{equation}{0}
\setcounter{figure}{0}
\setcounter{table}{0}
\setcounter{page}{1}
\makeatletter
\renewcommand{\theequation}{S\arabic{equation}}
\renewcommand{\thefigure}{S\arabic{figure}}
\renewcommand{\thetable}{S\arabic{table}}
\renewcommand{\bibnumfmt}[1]{[S#1]}
\renewcommand{\citenumfont}[1]{S#1}

\hspace{-22pt} \rule{1.0\textwidth}{2.5pt}

\begin{center}
	\Large{\textbf{An Alternative Strategy to Control the Electronic Properties of Bilayer Graphene: Semi-metal to Metal Transition and a 2D Material with Dirac Cone}} \\[0.5cm]
	
	\normalsize
%    \vspace{10pt}
%	\small

%\author{Srimanta Pakhira}
%\affiliation{Condensed Matter Theory, National High Magnetic Field Laboratory, Scientific Computing Department, Materials Science and Engineering, Florida State University (FSU), Tallahassee, Florida, 32310, USA}
%\affiliation{Department of Chemical \& Biomedical Engineering, FAMU-FSU Joint College of Engineering, Tallahassee, Florida, 32310, USA.}
%\author{Kevin P. Lucht}%
%\author{Jose L. Mendoza-Cortes}
%\email{Corresponding Author: Prof. J. L. Mendoza-Cortes, email: mendoza@eng.famu.fsu.edu\\[-3pt]} 
%
%\affiliation{Condensed Matter Theory, National High Magnetic Field Laboratory, Scientific Computing Department, Materials Science and Engineering, Florida State University (FSU), Tallahassee, Florida, 32310, USA}
%\affiliation{Department of Chemical \& Biomedical Engineering, FAMU-FSU Joint College of Engineering, Tallahassee, Florida, 32310, USA.}

	Srimanta Pakhira$^{1, 2}$, Kevin P. Lucht$^{1, 2}$, and Jose L. Mendoza-Cort\'es$^{1, 2, *}$\\
	\vspace{5pt}

	$^{1}$ Condensed Matter Theory, National High Magnetic Field Laboratory, Scientific Computing Department, Materials Science and Engineering, Florida State University (FSU), Tallahassee, Florida, 32310, USA\\
	
	$^{2}$ Department of Chemical \& Biomedical Engineering, FAMU-FSU Joint College of Engineering, Tallahassee, Florida, 32310, USA.\\
	
	\vspace{10pt}
	\normalsize
    E-mail:
	\href{mailto:mendoza@eng.famu.fsu.edu}{mendoza@eng.famu.fsu.edu}
\end{center}

\vspace{10pt}

\hspace{-23pt}
\rule{1.0\textwidth}{2.5pt}

\vspace{13pt}

\normalsize
\twocolumngrid

\normalsize
\section{I. Optimized geometry, band structure and density of states of AA-stacked bilayer graphene}

\begin{figure}[h!]
	\vspace{-8mm}
	\centering
	\includegraphics[width=0.98\linewidth]{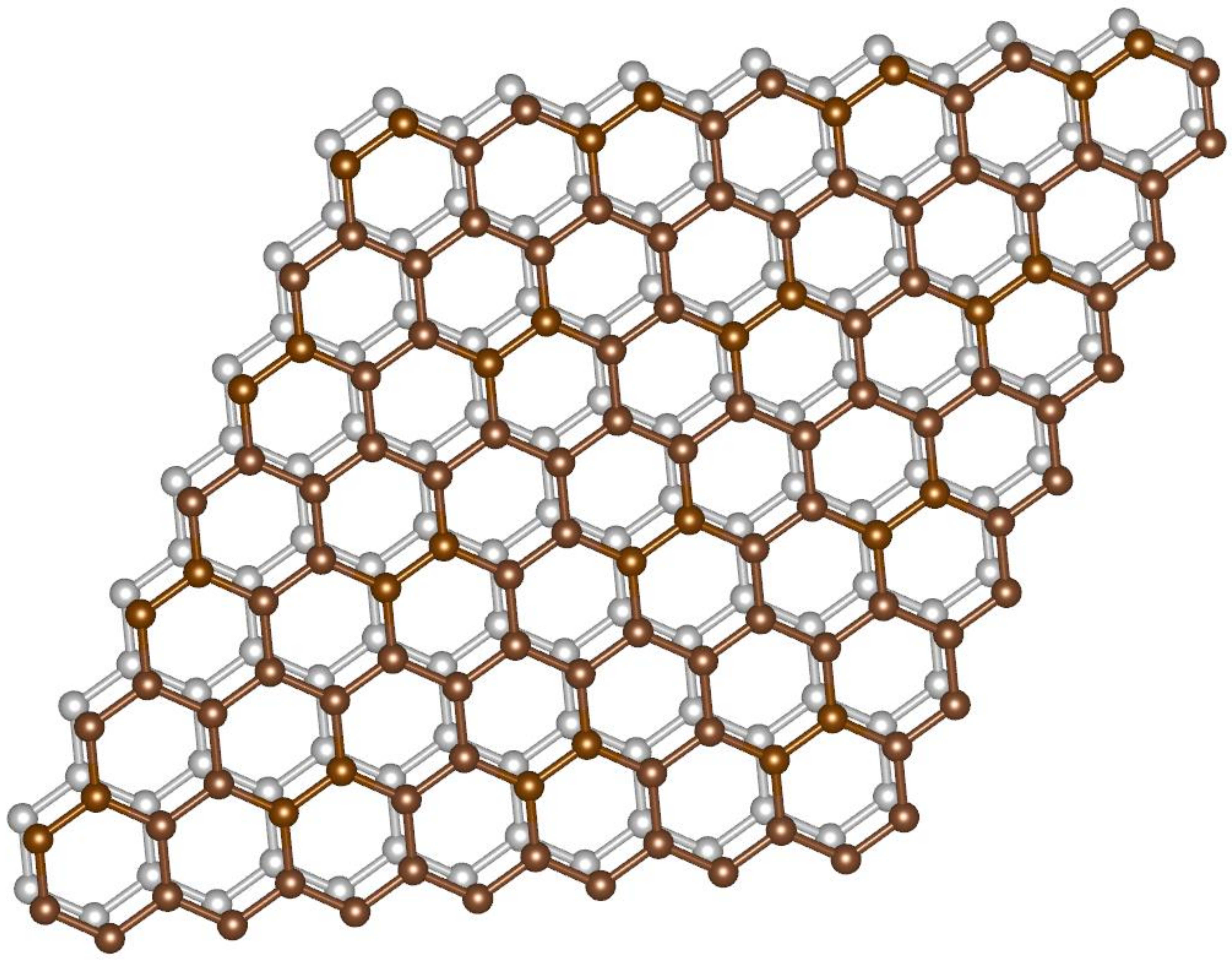}
	\caption{The optimized structures of AA-stacked bilayer graphene (BLG) 2D materials.}
	\label{fig SM1:AA-Stacked-Structure}

\end{figure}

\begin{figure}[h!]
	\centering
	\includegraphics[width=0.98\linewidth]{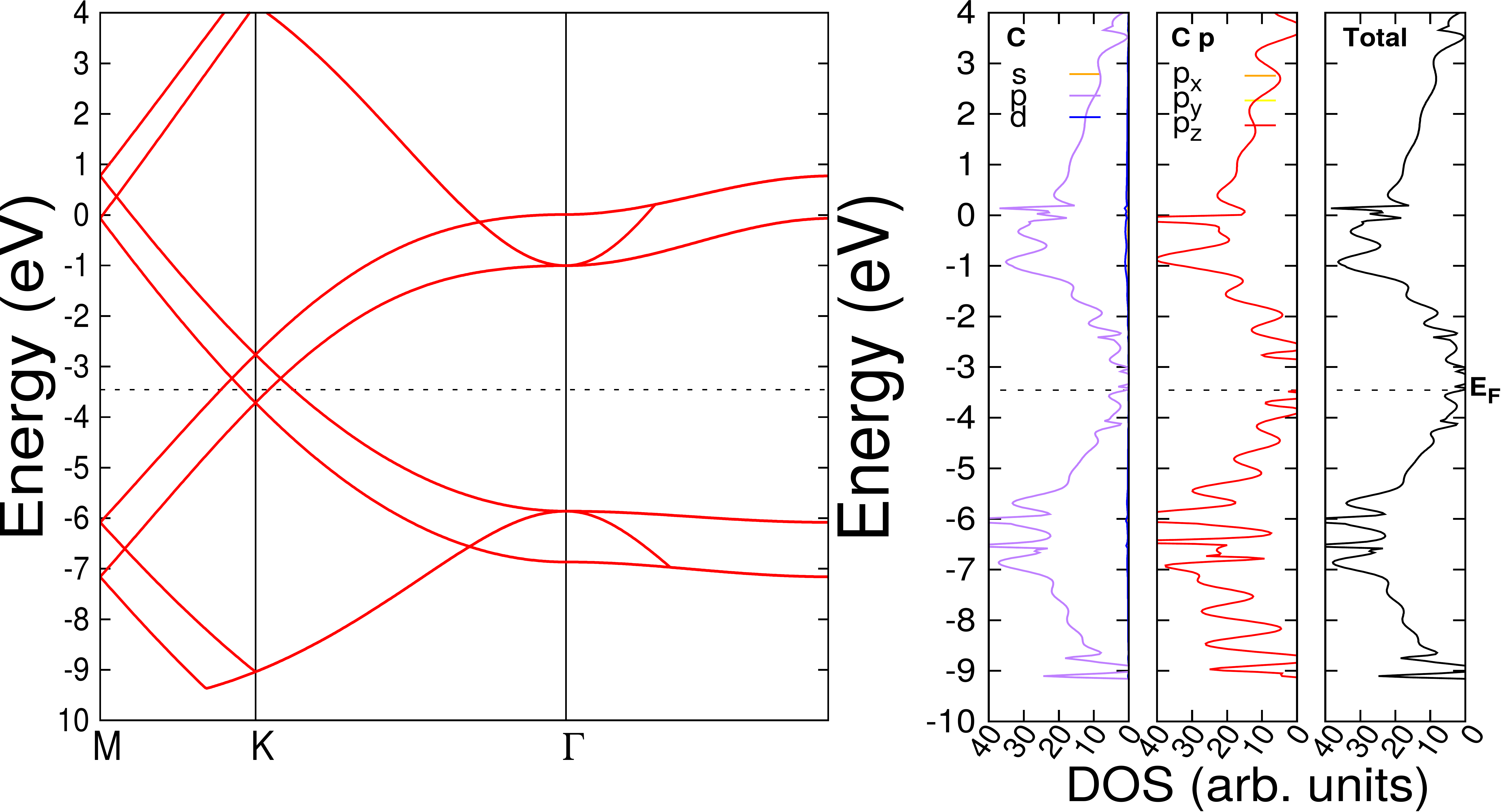}
	\caption{The band structure and density of states (DOSs) of AA-stacked BLG. The individual components of DOSs of the C atom, and total DOSs (depicted by ``Total'' in the third row) are also presented in here.}
	\label{fig SM2:AA-Stacked-BLG-DOSS_DPDOSS_BAND-2D}
	\vspace{-4mm}
\end{figure}

The optimized structure of AA-stacked bilayer graphene (BLG) is shown in Figure \ref{fig SM1:AA-Stacked-Structure}, and the band and DOSs are shown in Figure \ref{fig SM2:AA-Stacked-BLG-DOSS_DPDOSS_BAND-2D}. The band and DOSs calculations reveal that AA-stacked BLG is a non-zero band gap semiconductor, and it has an indirect band gap around 0.25 eV as depicted in the band structure. The individual components of p-orbital (i.e. p$_x$, p$_y$, p$_z$ sub-shells of p-orbital) are calculated along with total DOSs. We found the p$_z$ sub-shell of p-orbital accounts for the largest electron contribution in total DOSs.  

\section{II. Effect of spin alignment in BLG-2V and BLG-3V materials}

An alternative method of controlling the electronic properties of BLG-intercalated materials is by altering the spin configuration. We carried out Mulliken spin density analysis to study the spin configuration. Thus the estimation of spin can vary using another method to estimate it from the wave function.  All previous materials have referenced Vanadium in the high-spin state, and in the case of BLG-2V and BLG-3V in a ferromagnetic (FM) spin arrangement. Considering an anti-ferromagnetic (AFM) arrangement of spins, we can drastically modify the properties. The spin conformations of the BLG-intercalated materials are reported in Table S1.  

\begin{table}[h!]
	\vspace{-3mm}
	\caption{ Different Mulliken spin densities of the BLG-intercalated materials.} 
	\label{table SMI:mytableI}
	\tabcolsep=5.0pt
	\begin{tabular}{ c c c  }
		\hline 
		Materials   &  Average Spin of V  &  Total Spin \\ 
		\hline 
		BLG-1V      &  2.245               &  2.082      \\
		FM BLG-2V   &  1.856               &  3.740      \\
		FM BLG-3V   &  1.489               &  4.375      \\
		AFM BLG-2V  &  0.000   	           &  0.000 	 \\
		AFM BLG-3V  &  0.157               &  0.405  	  \\ 
		
		\hline 
		\vspace{-6mm}
	\end{tabular} 
\end{table}

\begin{figure}[h!]
	\includegraphics[width=0.99\linewidth]{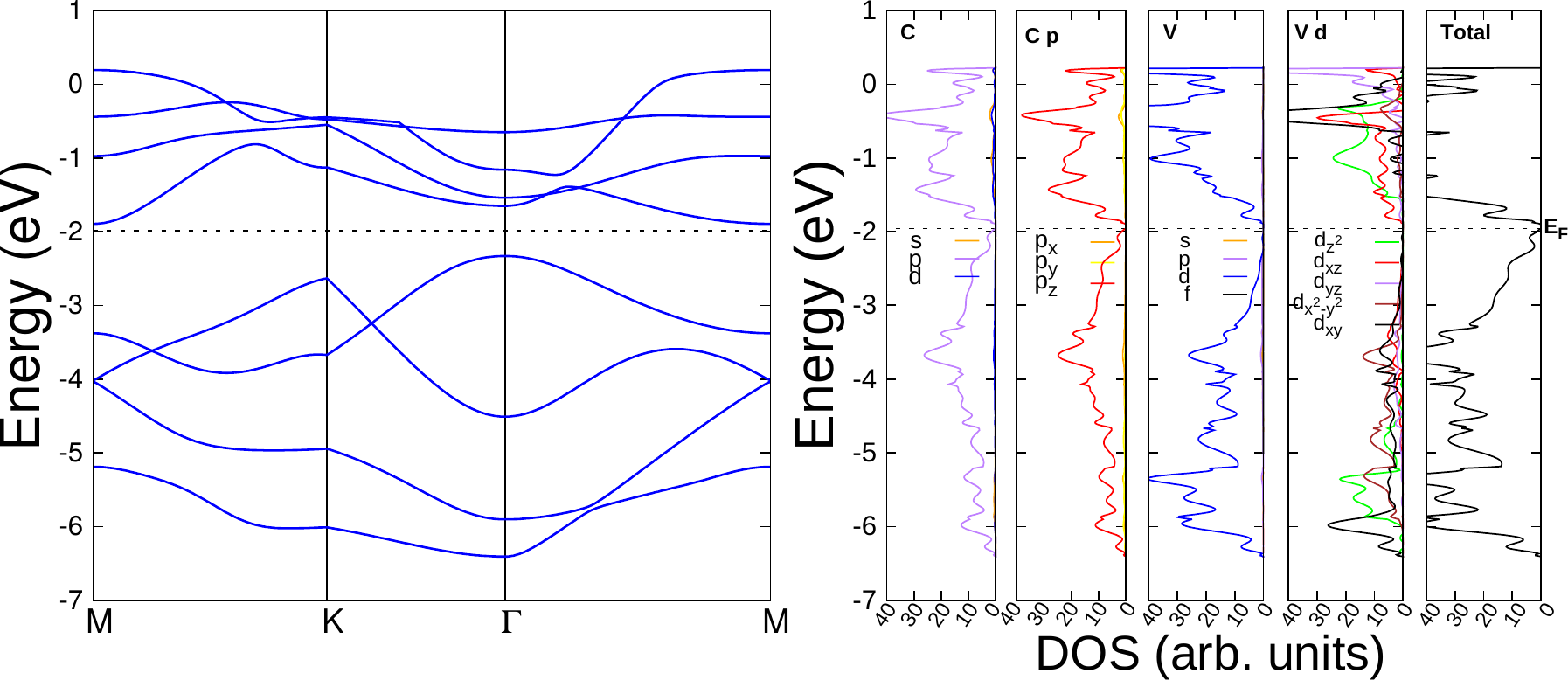}
	\vspace{-6mm}
	\caption{The band structure and DOSs of the alpha electron of AFM BLG-2V materials are presented. The beta electrons exhibit identical information. The individual components of DOSs of the C and V atoms, and total DOSs are also presented here. Compared to the FM, in the AFM state a band gap appears.}
	\label{fig:V2_graphene_AFM_2DPROP}
	\vspace{-3mm}
\end{figure}

Figure \ref{fig:V2_graphene_AFM_2DPROP} shows the band structure and DOSs of the AFM BLG-2V material. In the case of BLG-2V, the AFM conformation yielded higher spin of individual V atoms, 2.22 and -2.22 compared to the FM conformation with 1.86 for each Vanadium. Thus the average spin of AFM BLG-2V material is 0.0. Relative to its FM counterpart, the AFM BLG-2V structure is more stable by $\Delta G\!_{f}$ = -0.410 eV. Examining the electronic properties, we noticed considerable differences between the AFM and FM BLG-2V structures. Once in the AFM state, BLG-2V have a degenerate pair of band structures (for alpha and beta electrons) with a band opening between the valence and conductive band. The band gaps of this spacing are 0.101 eV for the indirect transition and 0.681 eV for the direct transition. Thus this calculation reveals that the band gap in BLG-intercalated material can be formed by altering the spin alignment.    

\begin{figure}[h!]
	\centering
	\includegraphics[width=0.99\linewidth]{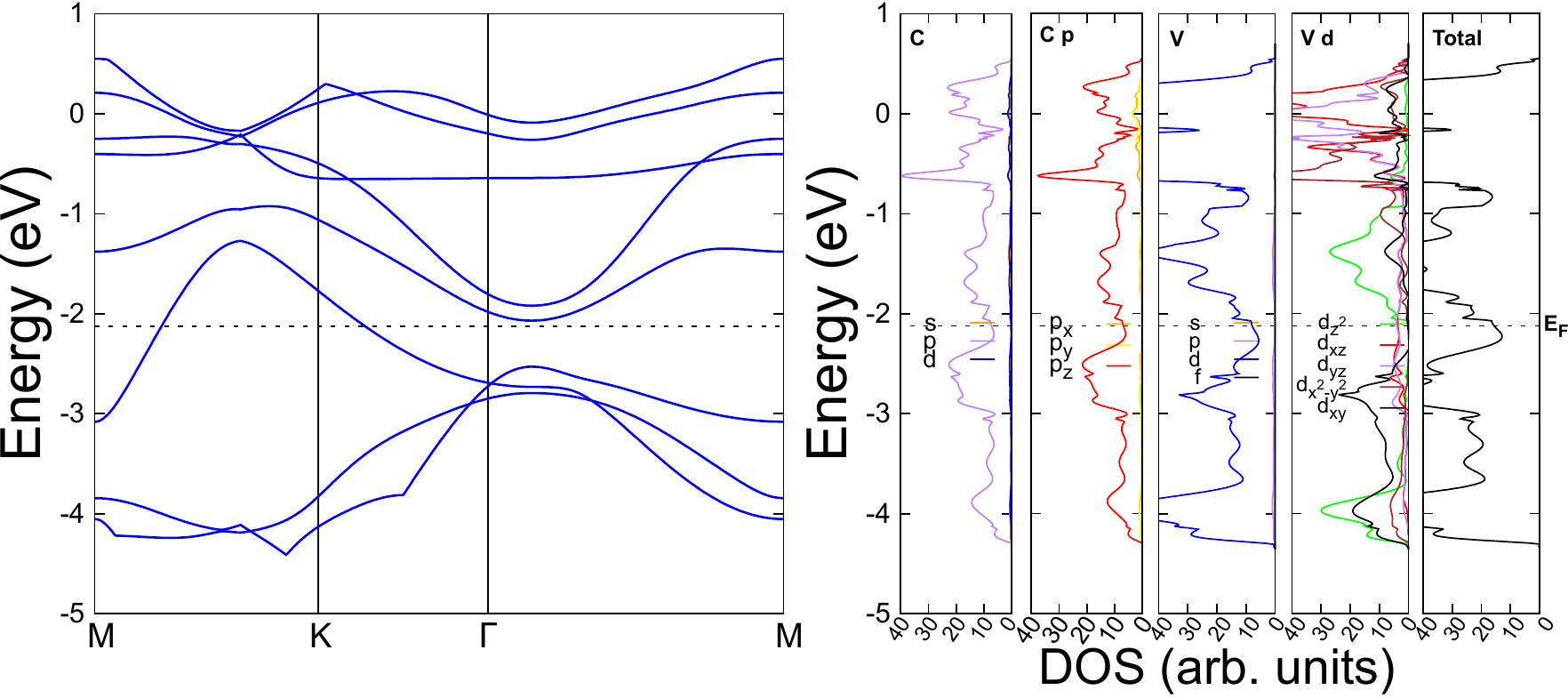}
	\vspace{-5mm}
	\caption{The band structure and DOSs of the alpha electron of AFM BLG-3V materials are presented. The individual components of the DOSs contributions from the C and V atoms, and total DOSs are also presented here. }
	\label{fig:V3_graphene_AFM_PROP}
	\vspace{-4mm}
\end{figure}

\begin{figure}[h!]
	\centering
	\includegraphics[width=0.99\linewidth]{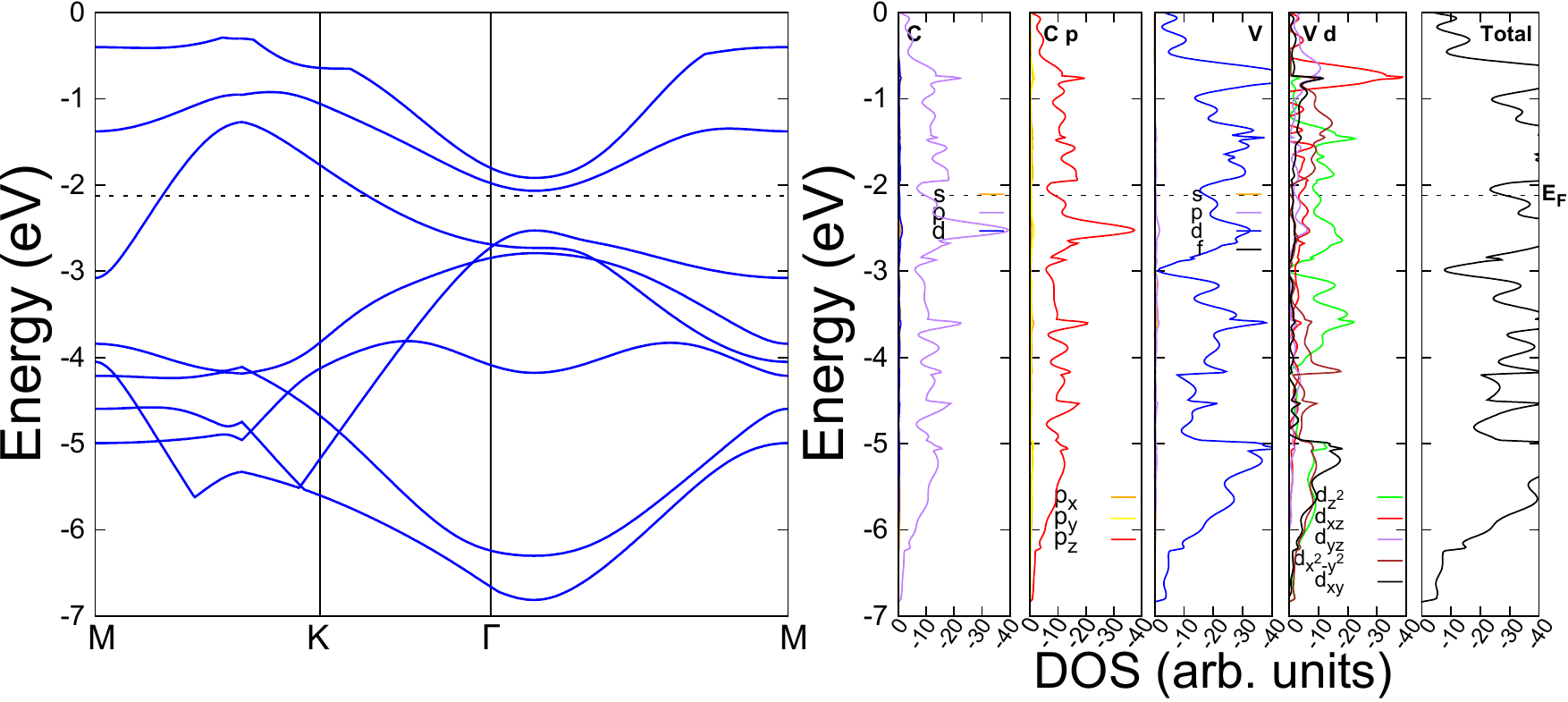}
	\vspace{-5mm}
	\caption{The band structure and DOSs of the beta electron of AFM BLG-3V materials are presented. The individual components of the DOSs contributions from the C and V atoms, and total DOSs are also presented here. Just as the alpha electrons, we can see the conductive behavior for the beta electrons.}
	\label{fig:V3_graphene_AFM_BETA_PROP}
	\vspace{-6mm}
\end{figure}

For the case of BLG-3V, there also exists one AFM conformation given the symmetry of the unit cell consisting of two Vanadium atoms with electrons in the alpha state and one in the beta state, or vice versa. We found that the AFM BLG-3V structure choses to allocate the spin as -1.726 for the beta state Vanadium, and the two alpha state Vanadiums with spins of 1.098. The average spin of AFM BLG-3V is 0.405. Comparing the structure to the FM BLG-3V structure, it is slightly less favorable by  $\Delta G\!_{f}$ = 0.193 eV. As for the electronic properties, we found for alpha and beta electrons that they are both highly conductive with a large DOSs and band overlap around the Fermi Energy (see Figure \ref{fig:V3_graphene_AFM_PROP} and Figure \ref{fig:V3_graphene_AFM_BETA_PROP}) resulting in a large electron density around the E$_F$. The present calculations showed that both the FM and AFM states of BLG-3V material are conducting and thus they show ordinary metallic behavior. This study of spin behavior shows that not only the electronic properties of BLG-intercalated materials dependent on the concentration of Vanadium atoms intercalated in BLG, but also on the arrangement of the spin, which adds an additional variable to modifying the behavior of BLG-intercalated materials.\\

\vspace{4mm}
Thus, we show an alternative method of controlling the electronic properties of BLG-intercalated materials is by altering the spin conformation. The AFM arrangement of spins in the BLG-2V material changes it from a metal to a semi-conductor with a small band gap. And for the BLG-3V material, the AFM state changes the electron density around the Fermi level compared FM states. We have observed that the 3d orbitals of V receive more electron donation from the graphene 2p$_z$ sub-shell when the total spin is increased. We have found the 2p$_z$ sub-shell of p-orbital of C atoms and 3d$_{yz}$ sub-shell of d-orbital of V atoms account for the largest electron contribution in total DOSs.\\

\section{All the .cif structures}
The optimized crystallographic information (.cif files) are provided below.

\vspace{2mm}
\subsection{Monolayer Graphene}
\vspace{-5mm}
\begin{lstlisting}
data_Monolayer_Graphene.out
_symmetry_space_group_name_H-M    'P6/MMM'
_symmetry_Int_Tables_number       191
_symmetry_cell_setting            hexagonal
loop_
_symmetry_equiv_pos_as_xyz
  'x,y,z'
_cell_length_a                    2.4515
_cell_length_b                    2.4515
_cell_length_c                    500.0000
_cell_angle_alpha                 90.0000
_cell_angle_beta                  90.0000
_cell_angle_gamma                 120.0000
loop_
_atom_site_label
_atom_site_type_symbol
_atom_site_fract_x
_atom_site_fract_y
_atom_site_fract_z
_atom_site_U_iso_or_equiv
_atom_site_adp_type
_atom_site_occupancy
C001   C     0.33333  -0.33333   0.00000   0.00000  Uiso   1.00


\end{lstlisting}

\vspace{2mm}
\subsection{AA-stacked Bilayer Graphene}
\vspace{-5mm}
\begin{lstlisting}
data_AA-Stacked_Bilayer_Graphene.out
_symmetry_space_group_name_H-M    'P6/MMM'
_symmetry_Int_Tables_number       191
_symmetry_cell_setting            hexagonal
loop_
_symmetry_equiv_pos_as_xyz
  'x,y,z'
_cell_length_a                    2.4490
_cell_length_b                    2.4490
_cell_length_c                    500.0000
_cell_angle_alpha                 90.0000
_cell_angle_beta                  90.0000
_cell_angle_gamma                 120.0000
loop_
_atom_site_label
_atom_site_type_symbol
_atom_site_fract_x
_atom_site_fract_y
_atom_site_fract_z
_atom_site_U_iso_or_equiv
_atom_site_adp_type
_atom_site_occupancy
C001   C    -0.66667  -0.33333   0.00320   0.00000  Uiso   1.00

\end{lstlisting}

\vspace{2mm}
\subsection{AB-stacked Bilayer Graphene}
\vspace{-5mm}
\begin{lstlisting}
data_AB-Stacked-Bilayer-Graphene.out
_symmetry_space_group_name_H-M    'P-3M1'
_symmetry_Int_Tables_number       164
_symmetry_cell_setting            trigonal
loop_
_symmetry_equiv_pos_as_xyz
  'x,y,z'
_cell_length_a                    2.4491
_cell_length_b                    2.4491
_cell_length_c                    500.0000
_cell_angle_alpha                 90.0000
_cell_angle_beta                  90.0000
_cell_angle_gamma                 120.0000
loop_
_atom_site_label
_atom_site_type_symbol
_atom_site_fract_x
_atom_site_fract_y
_atom_site_fract_z
_atom_site_U_iso_or_equiv
_atom_site_adp_type
_atom_site_occupancy
C001   C     0.00000   0.00000   0.00304   0.00000  Uiso   1.00
C003   C    -0.33333   0.33333   0.00304   0.00000  Uiso   1.00

\end{lstlisting}

\vspace{2mm}
\subsection{BLG-intercalated BLG-1V}
\vspace{-5mm}
\begin{lstlisting}
data_BG_V_graphene.out
_symmetry_space_group_name_H-M    'P6/MMM'
_symmetry_Int_Tables_number       191
_symmetry_cell_setting            hexagonal
loop_
_symmetry_equiv_pos_as_xyz
  'x,y,z'
_cell_length_a                    4.9423
_cell_length_b                    4.9423
_cell_length_c                    500.0000
_cell_angle_alpha                 90.0000
_cell_angle_beta                  90.0000
_cell_angle_gamma                 120.0000
loop_
_atom_site_label
_atom_site_type_symbol
_atom_site_fract_x
_atom_site_fract_y
_atom_site_fract_z
_atom_site_U_iso_or_equiv
_atom_site_adp_type
_atom_site_occupancy
C001   C    -0.33601  -0.16801   0.00344   0.00000  Uiso   1.00
C007   C     0.33333  -0.33333   0.00339   0.00000  Uiso   1.00
V009   V     0.00000   0.00000   0.00000   0.00000  Uiso   1.00

\end{lstlisting}

\vspace{2mm}
\subsection{BLG-intercalated BLG-2V}
\vspace{-5mm}
\begin{lstlisting}
data_BG_V2_graphene.out
_symmetry_space_group_name_H-M    'CMM2'
_symmetry_Int_Tables_number       35
_symmetry_cell_setting            orthorhombic
loop_
_symmetry_equiv_pos_as_xyz
  'x,y,z'
_cell_length_a                    5.0217
_cell_length_b                    8.5748
_cell_length_c                    500.0000
_cell_angle_alpha                 90.0000
_cell_angle_beta                  90.0000
_cell_angle_gamma                 90.0000
loop_
_atom_site_label
_atom_site_type_symbol
_atom_site_fract_x
_atom_site_fract_y
_atom_site_fract_z
_atom_site_U_iso_or_equiv
_atom_site_adp_type
_atom_site_occupancy
C001   C     0.24997   0.08385  -0.00854   0.00000  Uiso   1.00
C010   C     0.24996   0.08385  -0.00125   0.00000  Uiso   1.00
C005   C     0.00000  -0.16606  -0.00840   0.00000  Uiso   1.00
C007   C     0.00000  -0.33405  -0.00840   0.00000  Uiso   1.00
C014   C     0.00000  -0.16608  -0.00139   0.00000  Uiso   1.00
C016   C     0.00000  -0.33403  -0.00139   0.00000  Uiso   1.00
V009   V     0.00000   0.00000  -0.00490   0.00000  Uiso   1.00
V018   V     0.50000   0.00000  -0.00490   0.00000  Uiso   1.00

\end{lstlisting}

\vspace{2mm}
\subsection{BLG-intercalated BLG-3V}
\vspace{-5mm}
\begin{lstlisting}
data_BG_P6MMM_V3_graphene.out
_audit_creation_method            'Materials Studio'
_symmetry_space_group_name_H-M    'P6/MMM'
_symmetry_Int_Tables_number       191
_symmetry_cell_setting            hexagonal
loop_
_symmetry_equiv_pos_as_xyz
  'x,y,z'
_cell_length_a                    5.0319
_cell_length_b                    5.0319
_cell_length_c                    500.0000
_cell_angle_alpha                 90.0000
_cell_angle_beta                  90.0000
_cell_angle_gamma                 120.0000
loop_
_atom_site_label
_atom_site_type_symbol
_atom_site_fract_x
_atom_site_fract_y
_atom_site_fract_z
_atom_site_U_iso_or_equiv
_atom_site_adp_type
_atom_site_occupancy
C001   C     0.33575   0.16787   0.00359   0.00000  Uiso   1.00
C013   C    -0.33333   0.33333   0.00385   0.00000  Uiso   1.00
V017   V     0.50000  -0.00000   0.00000   0.00000  Uiso   1.00

\end{lstlisting}

\end{document}